\documentclass{aa}  

\usepackage{graphicx}
\usepackage{txfonts}
\begin{document} 

   \title{Disk Evolution Study Through Imaging of Nearby Young Stars
(DESTINYS): Scattered light detection of a possible disk wind in RY\,Tau} 


\titlerunning{Scattered light disk wind in RY\,Tau}
\authorrunning{Valeg\r{a}rd et al.}

   \author{P.-G. Valeg\r{a}rd\inst{1}
          ,
          C.Ginski\inst{2,1}
          ,
          C. Dominik\inst{1}
          ,
          J. Bae\inst{3}
          ,
          M. Benisty\inst{4}
          ,
          T. Birnstiel\inst{5,6}
          ,
          S. Facchini\inst{7}
          ,
          A. Garufi\inst{8}
          ,
          M. Hogerheijde\inst{1,2}
          ,
          R.\,G.~van~Holstein\inst{9}
          ,
          M. Langlois\inst{10}
          ,
          C. F. Manara\inst{11}
          ,
          P. Pinilla\inst{12,13}
          ,
          Ch. Rab\inst{5,14}
          ,
          \'A. Ribas\inst{9}
          ,
          L.B.F.M. Waters\inst{15,16}
          ,
          J. Williams\inst{17}}

   \institute{Anton Pannekoek Institute for Astronomy (API), University of Amsterdam, Science Park 904, 1098 XH Amsterdam, The Netherlands, \email{p.g.valegard@uva.nl}
   \and Leiden Observatory, Leiden University, Niels Bohrweg 2, NL-2333 CA Leiden, The Netherlands
   \and Department of Astronomy, University of Florida, Gainesville, FL 32611, USA
   \and University of Grenoble Alps, CNRS, IPAG, 38000 Grenoble, France
   \and University Observatory, Faculty of Physics, Ludwig-Maximilians-Universität München, Scheinerstr. 1, 81679 Munich, Germany
    \and Exzellenzcluster ORIGINS, Boltzmannstr. 2, D-85748 Garching, Germany
    \and Dipartimento di Fisica, Universit\`{a} degli Studi di Milano, Via Celoria 16, I-20133 Milano, Italy
    \and INAF, Osservatorio Astrofisico di Arcetri, Largo Enrico Fermi 5, I-50125 Firenze, Italy
   \and European Southern Observatory, Alonso de C\'{o}rdova 3107, Casilla 19001, Vitacura, Santiago, Chile
   \and Centre de Recherche Astrophysique de Lyon, CNRS, UCBL, ENS Lyon, UMR 5574, F-69230, Saint-Genis-Laval, France 
   \and European Southern Observatory, Karl-Schwarzschild-Strasse 2, 85748 Garching bei M\"{u}nchen, Germany
    \and Max-Planck-Institut f\"{u}r Astronomie, K\"{o}nigstuhl 17, 69117, Heidelberg, Germany
    \and Mullard Space Science Laboratory, University College London, Holmbury St Mary, Dorking, Surrey RH5 6NT, UK 
    \and Max-Planck-Institut für extraterrestrische Physik, Giessenbachstrasse 1, 85748 Garching, Germany
    \and Institute for Mathematics, Astrophysics \& Particle Physics, Radboud University, P.O. Box 9010, MC 62, NL-6500 GL Nijmegen, the Netherlands
    \and SRON, Sorbonnelaan 2, 3484CA Utrecht, The Nederlands
    \and Institute for Astronomy, University of Hawaii, Honolulu, HI 96822, USA
    }

   \date{Received May 11, 2022; accepted Month date, 2022}

 
  \abstract
   {Disk winds are an important mechanism for accretion and disk evolution around young stars. The accreting intermediate-mass T-Tauri star RY Tau has an active jet and a previously known disk wind. Archival optical and new near-infrared observations of the RY Tau system show two horn-like components stretching out as a cone from RY Tau. Scattered light from the disk around RY Tau is visible in the near-infrared, but not seen at optical wavelengths. In the near-infrared, dark wedges separate the horns from the disk, indicating that we may see the scattered light from a disk wind.}
   {We aim to test the hypothesis that a dusty disk wind could be responsible for the optical effect in which the disk around RY Tau is hidden in the I band, but visible in the H band. This could be the first detection of a dusty disk wind in scattered light. We also want to constrain the grain size and dust mass in the wind and the wind-launching region.}
   {We used archived Atacama-Large-Millimetre-Array (ALMA) and Spectro-Polarimetric High-contrast Exoplanet REsearch (SPHERE) I band observations combined with newly acquired SPHERE H band observations and available literature to build a simple geometric model of the RY Tau disk and disk wind. We used Monte Carlo radiative transfer modelling \textit{MCMax3D} to create comparable synthetic observations that test the effect of a dusty wind on the optical effect in the observations. We constrained the grain size and dust mass needed in the disk wind to reproduce the effect from the observations.}
   {A model geometrically reminiscent of a dusty disk wind with small micron to sub-micron-sized grains elevated above the disk can reproduce the optical effect seen in the observations. The mass in the obscuring component of the wind has been constrained to $1\times10^{-9} M_{\odot} \leq M \leq 5\times10^{-8} M_{\odot}$, which corresponds to a mass-loss rate in the wind of about $\sim 1\times10^{-8}M_{\odot}\mathrm{yr}^{-1}$. }
   {A simple model of a disk wind with micron to sub-micron-sized grains elevated above the disk is able to prevent stellar radiation to scatter in the disk at optical wavelengths while allowing photons to reach the disk in the near-infrared. Estimates of mass-loss rate correspond to previously presented theoretical models and points towards the idea that a magneto-hydrodynamic-type wind is the more likely scenario.}

   \keywords{Protoplanetary disks --
                Radiative transfer --
                Stars: individual: RY Tau --
                Stars: winds, outflows
               }
   \maketitle

%
\section{Introduction}
The structure and evolution of protoplanetary disks (PPDs) is important for understanding planet formation. Spatially resolved observations of nearby young stars in scattered light at infrared and sub-millimetre wavelengths routinely observe disks with structures such as gaps, rings, and spiral arms \citep{2018ApJ...869L..41A, 2020A&A...633A..82G, 2022arXiv220309991B}. These structures are results of processes that ultimately lead to the dispersal of the disk within a few million years \citep{2007ApJ...667..308C, 2014prpl.conf..475A}. Many mechanisms have been proposed as explanations for the structures seen in the observations, including accretion, gravitational instability, planet formation, planet-disk interaction, photo-evaporation, and ice lines (see \citet{2018ApJ...869L..41A} for an overview). 


Photo-evaporative (PE) and magneto-hydrodynamic (MHD) disk winds are two mechanisms where material is launched from the disk surface. In the case of PE winds, the gas is heated by ionizing radiation from the star. The wind is then driven and accelerated by the thermal gradient, dragging along dust particles that are well-coupled to the gas. Depending on the type of ionizing radiation, the mass-loss rate in the disk varies from $\sim1\times10^{-10} M_{\odot} \mathrm{yr}^{-1}$ for extreme-ultraviolet (EUV) radiation \citep{1994ApJ...428..654H,2004ApJ...607..890F} to about $1\times10^{-8} M_{\odot} \mathrm{yr}^{-1}$ to $1\times10^{-7} M_{\odot} \mathrm{yr}^{-1}$  for far-ultraviolet (FUV) \citep{2004ApJ...611..360A, 2009ApJ...690.1539G} and X-ray radiation \citep{2009ApJ...699.1639E,2010MNRAS.401.1415O,2012MNRAS.422.1880O, 2019MNRAS.487..691P, 2021MNRAS.508.1675E, 2021MNRAS.508.3611P, 2021arXiv211010637F}. In numerical simulations, the grain sizes entrained in PE winds are typically $\sim$1-2 $\mathrm{\mu{}m}$ in radius \citep{2010MNRAS.401.1415O,2021MNRAS.502.1569B} with an upper limit of $\sim$10$\mathrm{\mu{}m}$ \citep{2020AA...635A..53F} for X-ray driven winds.

In the case of a magnetically driven wind, the presence of a weak magnetic field, together with a differentially rotating disk, creates a magnetic pressure gradient that drives the wind. The MHD wind extracts angular momentum from the disk when the gas follows outwards along the magnetic field lines, driving disk accretion. The launch mechanism is also thought to be responsible for high-velocity jet outflows \citep{2002ApJ...581..988C,2007A&A...469..811Z,2009MNRAS.400..820T,2012ApJ...757...65S,2013ApJ...774...12F}. Semi-analytical calculations of dust transport in MHD winds allow grains up to $1 \mathrm{\mu{}m}$ to be entrained in the MHD wind \citep{2019ApJ...882...33G}. Mass-loss rates of MHD winds are expected to be of the same order of magnitude as accretion rates \citep{2013ApJ...769...76B, 2022ApJ...926L..23H}. Tracing and diagnosing the disk winds could be the key to understanding disk evolution in the context of observed low turbulence values \citep{2015ApJ...813...99F, 2017ApJ...843..150F, 2018ApJ...856..117F, 2020ApJ...895..109F}. 

By using velocity shifts in emission lines of ionic, atomic, and molecular species, out-flowing material from young stars can be identified \citep{1987ApJ...321..473E}. Jet and micro-jet outflows can be traced in the high-velocity component of optical forbidden lines, while the slower disk wind can be traced in the low-velocity component of the line \citep{1995ApJ...452..736H, 2013ApJ...772...60R, 2014A&A...569A...5N}. Winds can also be studied using molecular lines such as [NeII] \citep{2007prpl.conf..277P,2012A&A...538A...2P} and CO \citep{2013A&A...555A..73K}.

The presence of an outflow from RY Tau is well known from infrared spectroscopy \citep{1991ApJ...367..173G},  combined spectroscopic and photometric observations \citep{2016AstL...42..193B, 2019ApJ...886..115Y}, high-resolution optical and infrared spectra \citep{2020A&A...643A..32G}, and high-resolution M-band spectroscopy \citep{2022AJ....163..174B}. \citet{2022AJ....163..174B} observe a high- and a low-velocity component in the M-band spectra and interpret this as a high-velocity jet and a low-velocity disk wind originating from the inner disk. This opens up the possibility that the dust producing this scattered light signature may be coming from a dusty disk wind instead of the remains of an infalling envelope. Dust coupled to the gas in the disk winds could indeed be observed in scattered light at optical and near-infrared wavelengths. By combining archival observations by SPHERE/ZIMPOL (Zurich IMaging POLarimeter; \citealt{Beuzit2019,Schmid2018}), sub-millimetre observations by Atacama Large Millimetre Array (ALMA), and newly acquired near-infrared observations  by SPHERE/IRDIS (Infra-Red Dual-band Imager and Spectrograph; \citealt{Dohlen2008}), we discuss the possible scattered light detection of the RY Tau disk wind.




The paper is structured in the following way: In Section 2 we present the stellar properties of RY Tau. We subsequently describe the observations, including both the archival data and the new observations, in Sections 3 and 4. We perform measurements and analyse our data in preparation for the disk model in Section 5, and the radiative transfer modelling is described in Section 6. We go on to discuss our result in Section 7 and summarize our conclusions in Section 8.  

\section{RY Tau, stellar properties}

RY Tau is an intermediate mass T-Tauri star with a surface temperature of $T_{\rm eff}=5945K \pm{142.5}$ \citep{2004AJ....128.1294C}. Using archival photometry, \citet{2021AA...652A.133V} determined the luminosity to be $L=11.97 L_{\odot}$, the mass $M=1.95 M_{\odot}$, and the age $~4$\,Myr by fitting a Kuruz stellar model to photometry based on the \textit{Hipparcos} distance. RY Tau is in the late stage of envelope dispersal and is surrounded both by an associated extended nebulosity as well as remains of the envelope close to the star \citep{1995AJ....109.1181N,2013ApJ...772..145T}. RY Tau is optically variable with occasional dimming similar to the UX Ori type variable stars where material in the inner disk leads to variable extinction \citep{2016AstL...42..193B}. It is also a strong variable X-ray source \citep{2016ApJ...826...84S} with an optically bright jet \citep{1990ApJ...354..687C,2008AA...478..779S,2009A&A...493.1029A,2015ApJ...804....2C,2018AA...619A...9S,2019AA...628A..68G}.


The distance estimate from \textit{Gaia Early Data Release 3 (EDR3)} is $\sim138 \mathrm{pc}$ \citep{2016A&A...595A...1G, 2021A&A...649A...1G}. The quality of the parallax, however, is not optimal due to surrounding associated nebulosity\footnote{See technical note GAIA-C3-TN-LU-LL-124-01 in the \textit{Gaia DR2} release documentation.}. Comparing this with \textit{Hipparcos} puts RY Tau at a distance of 133 pc and the proper motion further strengthens its membership in the Taurus star-forming region ($\sim140 \mathrm{pc}$) \citep{2019AA...628A..68G}. In this work, we use the \textit{Hipparcos} distance for RY Tau.

\section{Observations \& data reduction}

\subsection{SPHERE/IRDIS}

RY\,Tau was observed with SPHERE/IRDIS in dual-beam polarimetric imaging (DPI; \citealt{deBoer2020,vanHolstein2020}) on 17 December 2019. The observations were carried out in the H band with a coronagraphic mask of radius 92.5\,mas (\citealt{Carbillet2011,Guerri2011}), centred on the photo-centre of the primary star.
The observations consisted of 17 polarimetric cycles, each containing four exposures taken at half-wave-plate switch angles $0^\circ$, $45^\circ$, $22.5^\circ$, and $67.5^\circ$. The individual frame exposure time was set to 32\,s, resulting in a total integration time of 36.3\,min. The average seeing during the observations was 0.88\arcsec{} and the coherence time of the atmosphere was 5.8\,ms.\\
The data were reduced using the publicly available IRDIS Data reduction for Accurate Polarimetry (IRDAP) pipeline (\citealt{vanHolstein2020}). IRDAP performed image cleaning with static flat-field frames and a bad-pixel mask and subtracted sky calibration frames, taken as part of the RY\,Tau observation sequence. Images were then re-centred on the stellar position, using dedicated centre calibration frames, which show symmetric spots around the stellar position and which were introduced by the deformable mirror of the adaptive-optics (AO) system. Subsequently, IRDAP performed polarimetric differential imaging (\citealt{Kuhn2001}) to subtract the stellar light from the images, while retaining the polarized light that scattered from the circumstellar disk.
IRDAP then used its built-in Mueller-matrix model to correct the data for instrumental polarization and polarization crosstalk. Finally, it computed images of the azimuthal Stokes parameters $Q_\phi$ and $U_\phi$ (\citealt{Schmid2006}), following the definitions of \cite{deBoer2020}. The remaining stellar polarization halo was subtracted following the approach by \cite{Canovas2011}, using a measuring annulus centred on the bright speckle pattern at the AO correction radius. The resulting image of the disk is shown in Fig. ~\ref{fig:rytau-compilation}.

\subsection{Archival \textit{SPHERE/ZIMPOL}}

The observations in the I' band ($\lambda_c$=790nm) were published in \citet{2019AA...628A..68G}. We used their processed data and only briefly summarize the observation and data reductions strategy.\\ 
The observations were performed with the SPHERE sub-instrument ZIMPOL in the differential polarimetric imaging mode, alternating between two de-rotator orientations with a 60$^{\circ}$ difference and additional dithering. 
The observation sequence consisted of alternating slow and fast polarimetry mode observations. The former allowed for long individual exposures with a high signal-to-noise ratio in the outer disk region, while the latter aimed to resolve the innermost disk region, with short exposures. For the slow-polarimetric observations, a Lyot coronograph with a diameter of 155 mas was inserted into the beam to mask the central star. Individual frame integration times of 50s and 20s were used for the slow- and fast-polarimetric modes, respectively. 
The standard reduction used in \citet{2014ApJ...790...56A} and \citet{2016AA...588A...8G} was used to process the DPI observations to obtain the final Stokes parameters Q and U as well as the polarized intensity PI and the azimuthal Stokes parameters $Q_\phi$ and $U_\phi$. 
In this work, we only make use of the $Q_\phi$ image.


\subsection{Archival ALMA}
The ALMA observations were taken in ALMA Band 6 ($\lambda_{c}=1.3$\,mm) and were published in \citet{2018ApJ...869...17L}. The image presented in this work comes from a data reduction by \citet{2020ApJ...892..111F}.

\subsection{Photometry}
The photometry used for fitting the spectral energy distribution (SED) was obtained from the literature and existing photometrical catalogues and archives. When multiple measurements were found for the same wavelength, the brightest measurement was used. The fluxes and references to the data are found in table \ref{tbl:photometry} of the appendix.

   
   
%

%

%

\section{RY Tau in observations}

At sub-millimetre wavelengths, the ALMA observations show a highly inclined disk where the disk extends out to $\sim70$ AU. The inner $\sim10$ AU has a very low flux, which is consistent with a cavity devoid of millimetre-sized grains. In the outer disk, a narrow gap can be seen at a distance of $\sim45$AU (Fig. \ref{fig:rytau-compilation} rightmost panel). 

The two scattered light images are strikingly different, even though the wavelength changes only by a factor of 2 from the I band to the H band. In the I band image (Fig. \ref{fig:rytau-compilation}, leftmost panel), two almost symmetric horn-like features are visible. The 'Northern Horn', [a], extends to the north-east and the 'Southern Horn', [b], extends to the east-south-east. Together they are forming a dark V shape, [c], in the direction of the disk spin axis (green arrow). No disk signal can be seen to be consistent with the orientation of the disk derived from the sub-millimetre image. 

In the H band image (Fig. \ref{fig:rytau-compilation}, middle panel), on the other hand, the top side, [1], of the disk surface is clearly visible, which is consistent with the sub-millimetre observation. The horns visible in the I band image are visible here as well, that is to say [2] and [3], but they look sharper and are less prominent. They still create the same dark V shape [4] along the spin axis of the disk (green arrow). Two dark wedges separate the disk from the horns, one in the north [5] and one in the south-east [6].

The completely different view of the disk at optical wavelengths compared to the near-infrared must be an optical effect created by material located above the disk. We argue that this effect, in combination with the horn features, could be the detection of a dusty disk wind in scattered light that absorbs photons from the star preventing them from reaching the disk surface at optical wavelengths while allowing the near-infrared photons through. In section 6 we explain how we built a model and tested our hypothesis using radiative transfer modelling.

\section{Measuring the horns}

   \begin{figure*}[ht]
        \includegraphics[width=\textwidth]{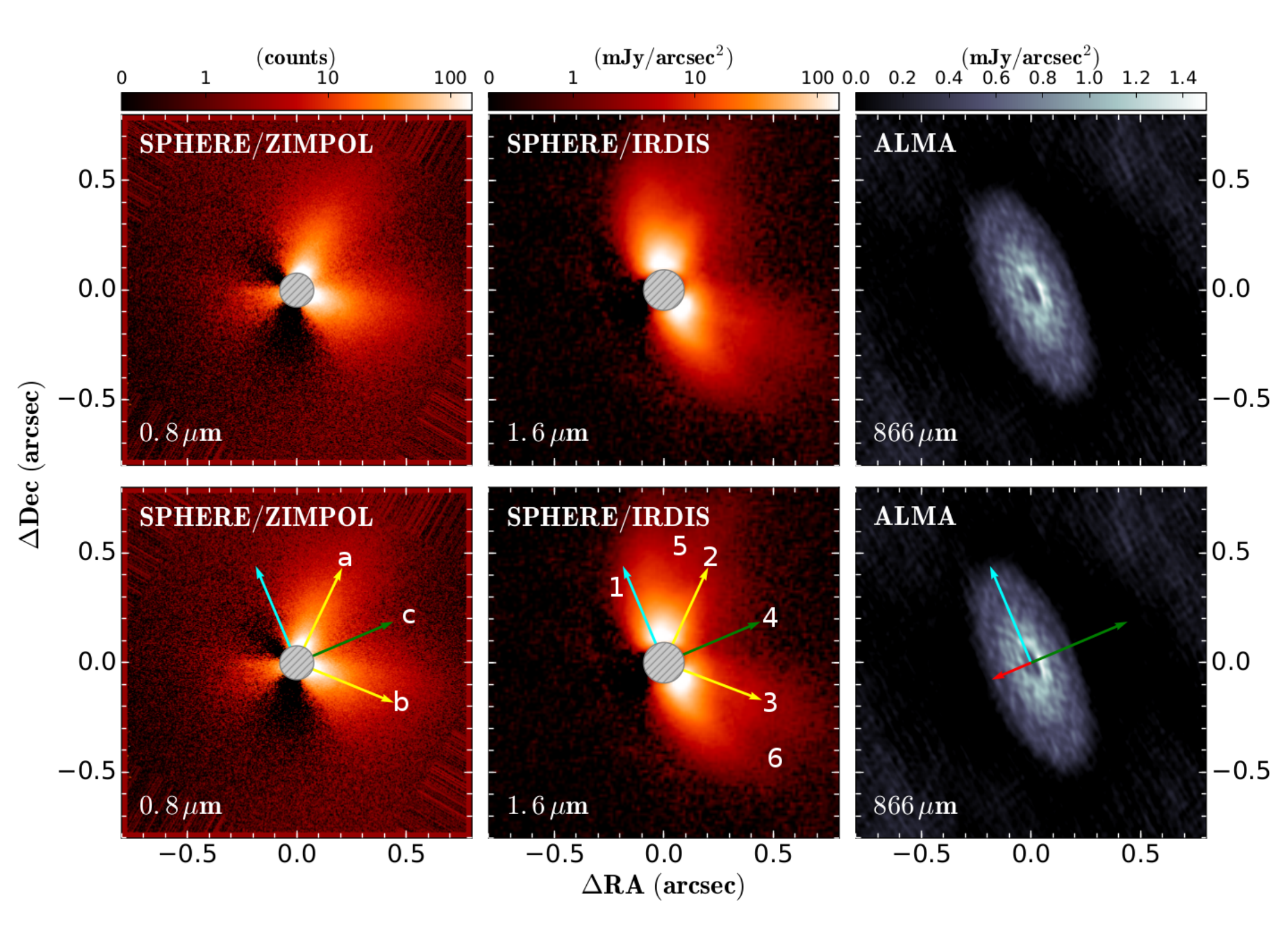}
        \caption{Images of RY Tau showing the observational data used in this paper. North is up and west is to the right. Leftmost panels: Observations by  the\textit{ SPHERE/ZIMPOL} I band. Middle panels: SPHERE/IRDIS H band. Rightmost panels: ALMA Band 6. Structures as described in section 4 for the I band: a) north-horn structure, b) south-horn structure, and c) disk normal. Structures as described in section 4 for  the H band: 1) disk-major axis, 2) north-horn structure, 3) south-horn structure, 4) disk normal, 5) north wedge, and 6) south wedge.} 
        \label{fig:rytau-compilation}
    \end{figure*}


The disk was observed at an inclination of i=65$^{\circ}$ and a position angle of 23$^{\circ}$ \citep{2018ApJ...869...17L}. To be consistent with scattered light images, however, we used the convention where a position angle was measured from the north anticlockwise where a 0 degree position angle is when the near-side of the disk is due west (see appendix Fig. \ref{fig:orientation}). With this convention, the P.A. became 203$^{\circ}$ and all subsequent angles were measured the same way. The outer edge of the disk was determined, by fitting an ellipse to the ALMA observation, to $\sim$ 70 AU. This is reasonable compared to R$_{\rm eff}$= 61 AU, which is the radius from where 90\% of the measured flux originates \citep{2018ApJ...869...17L}.


 
 Assuming that the brightest pixels trace the orientation of the horns, the angle of the horns was measured in the following way. For the ZIMPOL I band image, the two horns are located within 90 degrees of the disk normal. Therefore, annuli of a thickness of four pixels were created with increasing radius starting outside of the radius of the coronograph. The image was then rotated 23$^{\circ}$ so the disk normal pointed west and the near side of the disk pointed east. The maximum intensity was then found by finding the sum of the four horizontal pixels, due west, and bound by the annul, rotating the image 1$^{\circ}$ at the time, from 0 to 90$^{\circ}$ anticlockwise until the largest sum was found. The rotation that had the largest sum was then saved, and the radius of the annul was increased to cover the next four pixels. The same procedure was done clockwise between 0 and 90$^{\circ}$. This gave a rough estimate of the direction of the intensity maximum along the two horns. The procedure was then refined to an angle of $\pm 10^{\circ}$ around the found directions for each horn, and this time the rotation was only 0.1$^{\circ}$, but now seeking the brightest pixel. Measurements were done starting from the edge of the coronograph to the distance from the star when the pixel intensity dropped below a signal-to-noise ratio of three compared to the background, considering this radius to be the end of the horn.  
 The final angle of the horn was found by averaging the angles found for the brightest pixels per annul. The standard deviation for the values was then calculated.
 
 In the IRDIS H band image, the horns have a lower contrast to their surroundings due to the scattered light signal of the RY Tau disk. Therefore, the approach to find the horn direction in the IRDIS H band image needed to be slightly different to prevent picking up the signal from the disk, which lies within $\pm 90^{\circ}$ of the disk normal, and mistakenly identifying this with the wedge. We therefore lowered the upper limit of rotation to 65$^{\circ}$ in each direction, and in this way excluded the inner 15 pixels of the signal from the disk. These pixels were not used when calculating the average and standard deviation for the angles. Otherwise, we followed the same procedure as for the ZIMPOL measurement.
 
 The orientation of the horns anticlockwise from north, seen in the ZIMPOL I band image (Fig. \ref{fig:rytau-compilation}, yellow arrows), were found to be $338^{\circ}$ for the Northern Horn and $243^{\circ}$ for the Southern Horn with a standard deviation of $\sigma=3.88$ for the Northern Horn and $\sigma=1.84$ for the Southern Horn, respectively. The orientation of the horns seen in the IRDIS H band image (Fig. \ref{fig:rytau-compilation}, yellow arrows) were found to be $338^{\circ}$ for the Northern Horn and $245^{\circ}$ for the Southern Horn, with a $\sigma=2.88$ for the Northern Horn and $\sigma=4.93$ for the Southern Horn, respectively.

\section{Disk modelling}

To model the disk observations, we used the Monte Carlo radiative transfer code \textit{MCMax3D} \citep{2009A&A...497..155M}. In this section we begin by describing our model and then explaining our choices for the disk parameters used.

\subsection{Disk \& disk-wind model}

    \begin{figure}[ht]
        \includegraphics[width=8.7cm]{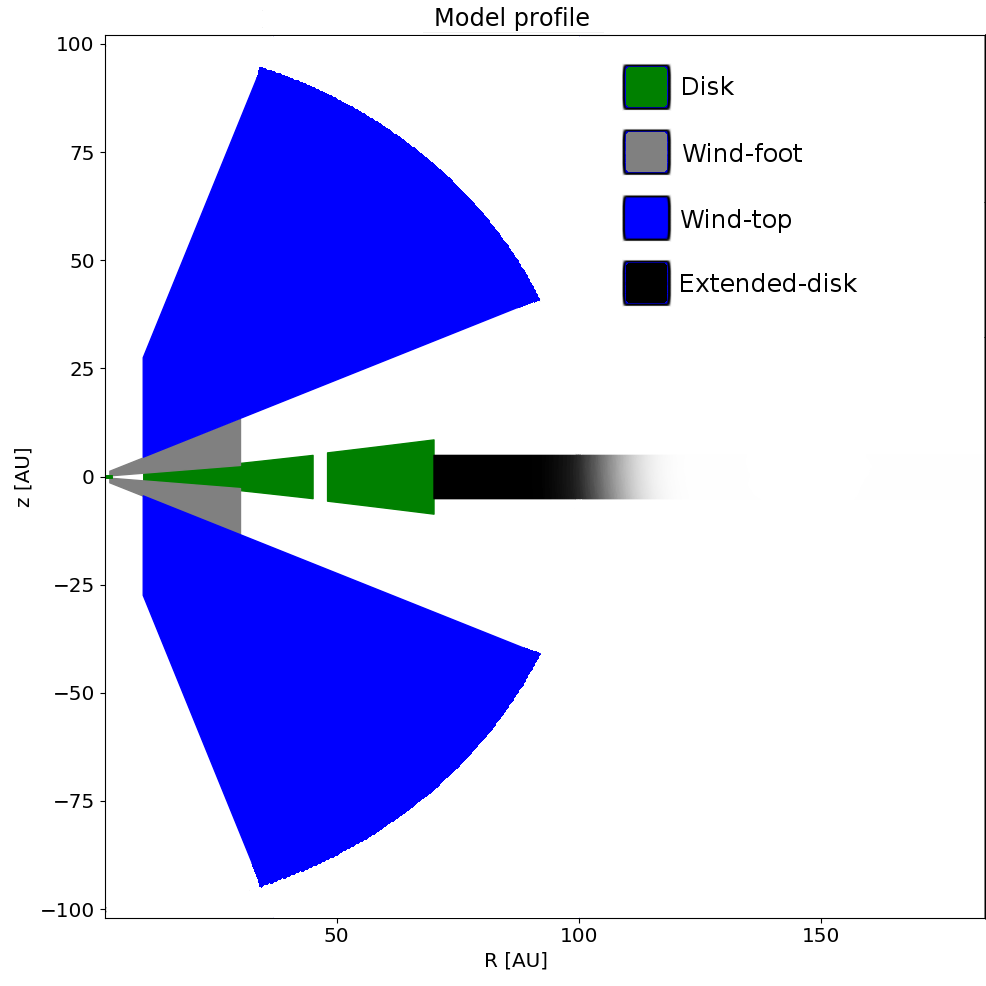}
        \caption{Model disk profile as seen in the plane of the disk. The disk (green), the wind foot (grey), the expanding wind top (blue), and the extended disk (black).}
        \label{fig:profile}
    \end{figure}

The geometry of the model we used is as follows (Fig. \ref{fig:profile}). The protoplanetary disk was modelled in three components (green), an inner ring close to the star ($0.2\leq R\leq3$ AU) and an outer disk ($10\leq R\leq 70$ AU) separated with a narrow gap ($44 \leq R\leq 47$ AU). The disk wind was modelled using two components, a wind foot (grey), which corresponds to the wind-launching region that shadows the disk at optical wavelengths, and a wind top (blue) of the freely expanding wind creating the horns seen in the observations. Outside the disk that is visible in scattered light, we assume that the disk still continues with the effect of blocking scattered light from the bottom side of the disk.

\subsection{Disks properties}

Since the modelling was performed in a multi-parameter space, we chose to fix some parameter values based on the existing observations and literature. Here we justify our choices for these parameters which can be found in table \ref{tab:mod_par_wind}. 

\begin{figure}[ht]
        \includegraphics[width=8.8cm]{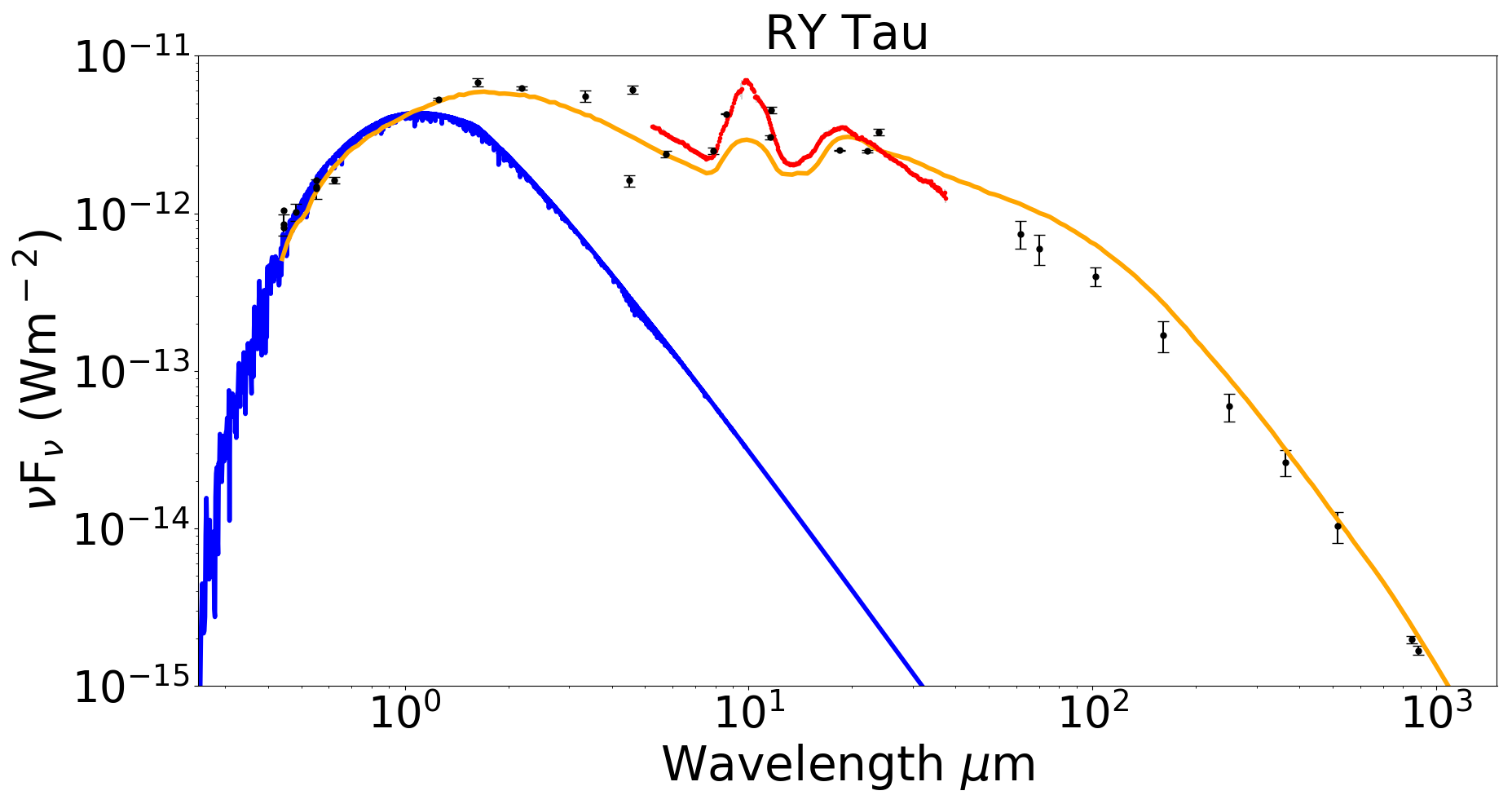}
        \caption{SED of RY Tau model (orange). The Kurucz model (blue) was fitted in \citet{2021AA...652A.133V} and the \textit{SPITZER} spectra (red) is from the \textit{CASSIS} database (AORkey 26141184).}
        \label{fig:rytau-wdiskwind}
    \end{figure}

The general size and extent of the disk model is based on the 1.3\,mm ALMA observation. It shows a disk with an outer edge of $\sim$ 70 AU with an inner 10 AU depletion of dust. The disk is separated by a gap at around 46 AU, which is shown by a drop in millimetre intensity. While the ALMA observations suggest inner disk depletion of millimetre-sized dust grains, the low resolution Spitzer spectra show a prominent narrow $10\, \mathrm{\mu{}m}$ silicate emission feature, indicating the presence of warm, small, inter stellar medium (ISM)-like grains. The near-infrared excess suggests there must be a dust component inside the 10 AU cavity that is not resolved in the ALMA observation. Its spatial size must either be lower than the resolution of the ALMA observation, or the dust mass in this region must be too small to register in the ALMA observation. The inner edge of this region is constrained by K-band interferometry and must be $r_{\rm inner}=0.21$ AU \citep{2020ApJ...897...31D}. Radiative-transfer modelling from \citet{2013ApJ...772..145T} shows that a scale height of 5AU at a radius of 50AU matches their observations and we, not being able to estimate the scale height from our observations, chose a similar scale height and a modest flaring $\beta=1.14$. Initially, we used gas-mass estimates for the disk mass calculated from the 1.3\,mm photometry range from $5\times10^{-2} M_{\odot}$ \citep{2019ApJ...872..158A} to $2\times10^{-4} M_{\odot}$ \citep{2011AA...535A..99M}. By comparing the model SED to the photometry, we adjusted the disk mass so that it reasonably fit with the millimetre flux (see Fig. \ref{fig:rytau-wdiskwind} and table \ref{tab:mod_par_wind}). These values fit well with the literature estimates for the disk mass. To describe the dust grain properties, we applied the \textit{DIANA} opacities \citep{2016AA...586A.103W} and set a gas-to-dust ratio of $100$ for all components. The dust density and settling parameter was set to $\Sigma \propto r^{-1} $ and $\alpha = 1\times10^{-4}$, respectively \citep{2018ApJ...869L..46D}.


An outer disk component was added as a continuation of the disk from $r_{\rm in}$=70 AU to $r_{\rm out}$=400 AU with a smaller scale height than that of the flared outer zone component. This was motivated by the observation that the bottom side of the disk is not visible in the scattered light observations. This is consistent with the fact that the true extent of the disks, as measured by CO observations, is usually much larger than the size seen in scattered light or the ALMA continuum. 

\subsection{Properties of the disk wind component}

To model the wind, we added a modified cone-shaped component split into two regions (see Fig. \ref{fig:profile}). Firstly we used a component, the wind foot, just above the disk, that casts the I'-band shadow on the disk, and secondly we used a component, the wind top, that allows for the shaping of the outflows visible as the two horns in the I'- and H-band observation. These two components have the same grain properties, consisting of small ISM-like grains, $\leq 1 \mathrm{\mu{}m}$. As shown in the ZIMPOL observation, the extent of this component is considerable, $\geq$ 100 AU. The wind foot has a radial extent from the inner disk zone to 30 AU. The radial extent of the wind foot was varied between 10 AU up to the outer edge of the disk at 70AU. Theoretical modelling by \citet{2022A&A...659A..42R} for a disk around a solar-mass star indicates that 90$\%$ of the dust entrained in the wind is launched from the inner disk to about $\sim$ 30 AU, a value we also find reasonable in our simple model. The density profile was modelled as $\propto r^{-2}$, which is consistent with spherical expansion at a constant speed, for both components of the wind. No settling was applied to the wind foot nor the wind top (i.e. $\alpha = 1$). The dust mass of the wind foot was varied from $1\times10^{-9} M_{\odot}$ to $5\times10^{-7} M_{\odot}$ to find the lower limit for when the disk became visible in the I band and a higher limit for when the disk disappeared in the H band. A best-fitting mass for the disk was determined by the mass when the disk disappeared in the I band.

\begin{table*}
\caption{Parameters of the model where the disk became obscured by the wind in the I band and while still visible in the H band.}\label{tab:mod_par_wind}
\centering
\renewcommand{\arraystretch}{1.5}
\begin{tabular}{lllllll}
\hline \hline
Parameter & Inner & Mid-zone & Outer & Extended & Disk wind & Disk wind \\
 & Zone & & Zone & Zone & foot & horns \\
\hline
$r_{\rm in}$ [AU]& $0.21$ & $10$ & $47$ & $70$ & $0.21$ & $10$ \\
$r_{\rm out}$ [AU]& $3$ & $44$ & $70$ & $400$ & $30$ & $100$\\
$M_{\rm dust}$ [$M_{\odot}$]& $1\times10^{-8}$ & $2\times10^{-4}$ & $1.4\times10^{-4}$ & $1\times10^{-5}$ & $1\times10^{-8}$ & $1\times10^{-8}$\\
$\Sigma\propto$ & $r^{-1.0}$ & $r^{-1.0}$ & $r^{-1.0}$ & $r^{-1.0}$ & & \\
$\rho\propto$ & & & & & $r^{-2}$ & $r^{-2}$ \\ 
\hline
$\beta$ & $1.0$ & $1.14$ & $1.14$ & $0$ &  &\\
h [AU]& $0.04$ & $0.9$ & $5.58$ & $5$ & &\\
$R_{h}$ [AU] & $0.4$ & $10$ & $47$ & $70$ & &\\
\hline
$a_{\rm min}$ [$\mathrm{\mu{}m}$] & $0.05$ & $0.05$ & $0.01$ & $0.01$ &$0.01$ & $0.01$ \\
$a_{\rm max}$ [$\mathrm{\mu{}m}$] & $1.0$ & $100$ & $100$ & $2$ & $1$ & $1$ \\ 
$a_{\rm pow}$ [$\mathrm{\mu{}m}$] & $3.5$ & $3.0$ &$3.0$ & $3.0$ &$3.5$ & $3.5$ \\
\hline
$\theta$ & $65$ & $65$ & $65$ & $65$ & $65$ & $65$ \\
$\phi$ & $113$ & $113$ & $113$ & $113$ & $113$ & $113$\\
$M_{\rm gas}/M_{\rm dust}$ & $100$ & $100$ & $100$ & $100$ & $100$ & $100$ \\
$\alpha$ & $1\times10^{-4}$ & $1\times10^{-4}$ & $1\times10^{-4}$ & $1\times10^{-4}$ & $1.0$ & $1.0$\\

\hline
\end{tabular}
\end{table*}

    
    
   \begin{figure*}[ht]
        \centering
        \includegraphics[width=0.8\textwidth]{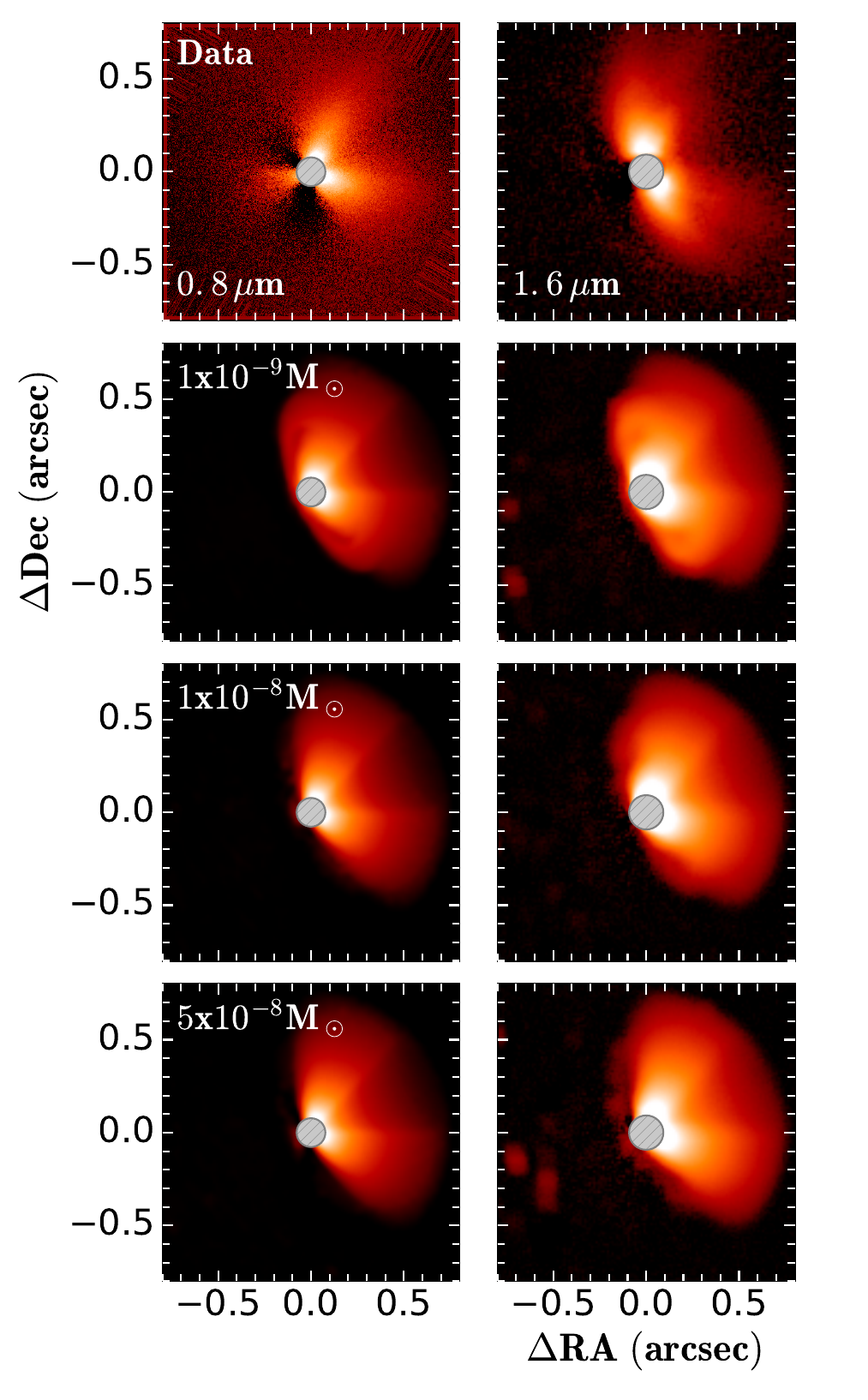}
        \caption{Comparison of data (first line) and models (one by line). }
        \label{fig:rytau-data-model}
    \end{figure*}

\section{Result \& discussion}

We find that the observations can be reproduced with a simple disk-wind-like model of small grains elevated above the disk, which can also explain the disappearance of the disk in the I' band while still being visible in the H band. Based on the I band image, \citet{2019AA...628A..68G} argued that the horn-like features belong to the protostellar envelope. The envelope, which could be a part of the nearby situated nebulosity, scatters the near-infrared photons from the much weaker disk, while allowing the stronger stellar photosphere to still be visible \citep{2019AA...628A..68G}. With the orientation of the jet and the dark V shape, the observation in the I band is reminiscent of an outflow cavity created by the jet in this protostellar envelope. 

With the addition of the new H band observations, we propose a different interpretation. The outflow from RY Tau is well known from spectroscopy \citep{1991ApJ...367..173G, 2016AstL...42..193B, 2019MNRAS.483..132P, 2019ApJ...886..115Y, 2020A&A...643A..32G}. Recent M-band spectroscopy of the ro-vibrational CO lines favours a 'disk + wind' interpretation, where the line of sight grazes the top of the dusty wind \citep{2022AJ....163..174B}. This means that the observer looks through the gas column of the wind towards the photosphere. The outflowing material creates a highly excited blue-shifted absorption feature (9-10 $\mathrm{km\,s}^{-1}$), suggesting the origin to be close to the star \citep{2022AJ....163..174B}. Since the disk is hidden in the I band, but visible at only slightly longer wavelengths (in the H band),  the size of the grains responsible for shadowing and hiding the disk must be small ($\leq 1 \mathrm{\mu{}m}$). We believe these grains shadowing the disk in the I band could be grains entrained in a MHD disk wind originating from the inner disk.    

The H band observation clearly suggests that elevated material is not present over the disk towards larger radii. We can see this in the dark wedges that separate the disk from the horns, both in the north and the west. If the elevated material is remains of the protostellar envelope, lit by the star, one would expect there to be no wedges between the disk and the envelope. The hypothesis of the morphology of a disk wind fits well with self-consistent modelling, involving a combination of MHD and photo-evaporation, by \citet{2022A&A...659A..42R}. They show that the majority ($\sim$ 90 $\%$) of the material is launched from the inner few tenths of AU which would produce a similar wedge effect. In our model, we are mimicking the wedge that separates the horns and the disk by introducing a two-component disk-wind region. Based on the inclinations measured for the horns and the scale height of the disk, we introduced a wind foot responsible for the shadowing of the disk and a wind top responsible for the visible horns. The radial extent of the wind foot could also be seen as the region from which the disk wind is launched.

The optical effect of the disappearing disk in the I' band in our model is due to the small grains in the launching region of the wind. The introduction of a wind-foot region with elevated small grains does, not only mask the stellar light from the disk in the I' band, but also -- together with the wind top -- generate the two 'horns' seen in both the I'- and H band observations (Fig. \ref{fig:rytau-data-model}). In the I band, the disk is visible for a wind-dust mass of the wind foot $1\times10^{-9} M_{\odot}$ and disappears as the mass increases above $1\times10^{-8} M_{\odot}$. The disk remains visible in the H band until $5\times10^{-8} M_{\odot}$. We therefore constrained the dust mass in the shadowing wind foot to $1\times10^{-9} M_{\odot} \leq M \leq 5\times10^{-8} M_{\odot}$. If no significant grain growth has taken place, all dust mass is contained in small grains and well mixed to the surface. In that case, a gas-to-dust ratio in the wind close to 100 is a reasonable assumption \citep{2022A&A...659A..42R}. Assuming a wind speed of $10\,\mathrm{km s}^{-1}$ \citep{2022AJ....163..174B} and the extent of the wind components in our models (100\,AU), this would correspond to an estimated lower limit for the mass-loss rate of $\sim 1\times10^{-8} M_{\odot}\mathrm{\mathrm{yr}}^{-1}$. If, on the other hand, there has been significant grain growth, then much of the grain mass is present in large grains which will be confined to the mid-plane and not available for entrainment into the wind and any attempt to entrain such larger grains would not work. In this case, the gas-to-dust ratio can be significantly larger than 100. Then the arrived dust masses that are required to produce the scattered light and optical depth effects would correspond to a significantly larger gas mass-loss rate.

Theoretical disk wind models predict a symmetric outflow in the case of disk winds \citep{ 2022A&A...657A..69F, 2022A&A...659A..90F, 2022A&A...659A..42R}, while synthetic observations by \citet{2020A&A...639A.137K} suggest an asymmetry in the case of infalling material. Observations of material colliding and interacting with the central star and disk in, for example, SU Aur \citep{2021ApJ...908L..25G} and GM Aur \citep{2021ApJS..257...19H} as well as WW Cha, J1615-1921, and DoAr 21 \citep{2020A&A...633A..82G}, do indeed show a lot of structure and asymmetries. The opening angle of the cone shape giving rise to the horns seen in the observations is bound by the grain size of the grains entrained in the wind. The measured semi-opening angle in the I band lobes and the H band horns ($\sim 43^{\circ}$) correspond to a grain size of $\sim 1-2 \mathrm{\mu{}m}$ for magnetically driven disk winds \citep{2022A&A...659A..42R}. For a magnetically launched wind, the launching angle corresponds to an expected semi-opening angle larger than $30^{\circ}$ \citep{1982MNRAS.199..883B}. Since we traced the horns using the brightest intensity, we matched the shape of the model to reproduce the horns in the direction of the brightest pixels. Thus, the spine of the horn features are within and in agreement with the models by \citet{1982MNRAS.199..883B}. However, this leads to the semi-opening angle of the wind-top geometry to be $\sim10^{\circ}$ smaller than this critical angle. This does not contradict the disk wind hypothesis, but means that the uppermost part of the wind top can still be shaped by the jet. The synthetic observations by \citet{2022A&A...659A..42R} also show material present above this critical angle for warm MHD winds. This is due to the fact that even if the launching angle does not exceed $60^{\circ}$ from the disk plane, the wind tends to collimate towards the vertical at larger scales (Dullemond personal communication). The close to symmetric horn structure in the observations of RY Tau, the opening angle and the presence of the dark wedges between the disk and the horns, all become strong indicators that the material observed above the disk is launched from the disk, rather than material falling onto the disk. 

The interferometric data \citep{2020ApJ...897...31D} and optical spectroscopy \citep{2019MNRAS.483..132P} suggest that the inner radius of the disk lies close to the star and that the inner disk is not completely empty. The 1.3 mm observation shows a cavity devoid of large grains, whereas the near-infrared flux in the SED is still high. The majority of dust in protoplanetary disks consists of silicate grains. The silicate emission seen in the \textit{SPITZER} low resolution spectra is very strongly peaked, and the shape also suggests that small grains are dominating the inner disk. In fact, RY Tau has one of the strongest and most peaked 10$\mathrm{\mu{}m}$ silicate emission features among the known IMTT stars \citep{2021AA...652A.133V}. In our model, however, the silicate feature is weaker than that of the observations by \textit{SPITZER} and not as strongly peaked. One would expect this material in the inner disk to be the source of silicate emission, but our simple model instead shows that the  small grains in the disk wind is the main contributor of 10 $\mathrm{\mu{}m}$ silicate emission. 





The final models contain some numerical noise in the H band images coming from stray photons that make it through the disk and scatter in the introduced outer disk region (see Fig. \ref{fig:rytau-data-model}). The appearance of these spots is due to low photon statistics in this region.

We compared our model and the observation with two recent theoretical papers of disk winds. One of the papers by \citet{2022A&A...657A..69F} models a wind fully driven by extreme-ultraviolet- (XEUV) driven photo-evaporation. The other paper by \citet{2022A&A...659A..42R} uses a non-ideal MHD model of the wind while also including XEUV heating to model PE flows. Both models assume a central star of $M_{\star}=0.7M_{\odot}$ with an X-ray luminosity of $L_{x}=2\times10^{30} \mathrm{erg s}^{-1}$. The PE models by \citet{2022A&A...657A..69F} assume that the inner disk is cleared of gas and dust to maximize the efficiency of photo-evaporation by using cavity sizes of 20 and 30\,AU, while in the MHD models by \citet{2022A&A...659A..42R} the cavity does not extend further than 2\,AU for most models. In comparison, RY Tau is more massive ($ 1.97M_{\odot}$) and has a variable X-ray flux that peaks at at least $\sim10^{31} \mathrm{erg s}^{-1}$ \citep{2016ApJ...826...84S}. One would expect this extra flux to increase the mass-loss rate in the wind from the disk for both models. 

However, the cavity in the disk of RY Tau in millimetre continuum is $\sim 10$ AU, that is to say smaller than in the best case assumed for the PE model. This means that the signal of a PE wind would likely be even lower than predicted by \cite{2022A&A...657A..69F} in the case of RY Tau.  In combination with a strong, $\mu\mathrm{m}$ silicate emission feature as well as significant near-infrared excess from material in the cavity (also strengthened by interferometry \citet{2020ApJ...897...31D}), the MHD model better mimics the observed geometry of the RY Tau system. One would therefore be inclined to prefer the MHD model over the PE model. This is not only supported from a geometrical point of view, but also since the active jet of RY Tau indicates a strong magnetic field which suggests that magneto-centrifugal disk winds are likely to be launched from within 1 AU \citep{2015ApJ...804....2C, 2019AA...628A..68G}. The opening angle between the horns could correspond to a magnetic wind having entrained dust particle sizes dominated by micron to sub-micron grains \citep{2022A&A...659A..42R}. This would be more compatible with the geometry of the observations and with our hypothesis. We therefore find it more likely that the wind-launching mechanism in this case is magneto-centrifugally driven.

The ratio of the mass-loss rate through a MHD wind compared to mass loss through accretion for the disk is expected to be $0.74 \lesssim \dot{M}_{\rm wind}/\dot{M}_{\rm acc} \lesssim 2.3$ \citep{2022ApJ...926L..23H}. For RY Tau, with an accretion rate of $6.4-9\times10^{-8}M_{\odot}\mathrm{yr}^{-1}$ \citep{2004AJ....128.1294C}, this would mean a mass-loss rate in the wind of  $4.7\times10^{-8} M_{\odot}\mathrm{yr}^{-1} \lesssim \dot{M}_{\rm wind} \lesssim 2.1\times10^{-7}M_{\odot}\mathrm{yr}^{-1}$, which corresponds fairly well with the mass-loss rate found with our simple model. \citet{2021A&A...652A..72A} measured the accretion rate of RY Tau in a more self-consistent way to $\sim3\times10^{-8}M_{\odot}\mathrm{yr}^{-1}$. Using the same method as above, this led to an estimate in a mass-loss rate of $2.22\times10^{-8} M_{\odot}\mathrm{yr}^{-1} \lesssim \dot{M}_{\rm wind} \lesssim6.9\times10^{-8}M_{\odot}\mathrm{yr}^{-1}$ which also corresponds well to our estimates for mass loss from the simple wind model. 

The conical morphology as observed by SPHERE is reminiscent of recently observed, conically shaped, molecular outflows in CO gas emission around DG Tau B \citep{2020AA...634L..12D} and HH30 \citep{2018A&A...618A.120L}. However, no spatially resolved gas-velocity data available for RY Tau show any similar structures. While the disk wind hypothesis is intriguing, our analysis does not yet rule out the possibility that the optically observed morphology around RY Tau comes from an outflow cavity as a result of the jet. However, we note that \citet{2022AJ....163..174B} detected a blue-shifted CO line in absorption of about $9-10 \mathrm{km s^{-1}}$, which they interpret as a disk wind signal originating from the inner disk which would be consistent with our favoured interpretation.

RY Tau could be the first source where the presence of a dusty disk wind has been observed in scattered light. It is unlikely that a PE disk wind would be detected in scattered light with SPHERE/IRDIS since the weak signal predicted for these models would not rise above the observational noise, which would make detection challenging. Photo-evaporative winds might be detectable using the JWST NIRCam \citep{2022A&A...657A..69F}. Disk winds should probably lie just at or below the detection limit for most systems \citep{2022A&A...657A..69F,2022A&A...659A..42R}. However, when viewed at a high inclination ($\sim65^{\circ}$), the wind of RY Tau might be more easily seen as the contrast between the disk and the wind which decreases in nearly edge-on observations \citep{2022AJ....163..174B}. Comparing with the observed disks by \citet{2020ApJ...892..111F}, it belongs to only $8\%$ of disks with inclinations above $60^{\circ}$. Considering its accretion rate, jet, X-ray luminosity, and inclination,  RY Tau is a rather unique system where several parameters could conspire to allow the disk wind to be detected.

\section{Conclusion}

We present a new interpretation of the scattered light observations of the system of RY Tau. Using a simple geometric model we show that the absence of a disk detection at optical wavelengths could be due to a dusty disk wind. Our conclusions are as follows:

\begin{enumerate}
    \item
    Small dust grains elevated above the disk are able to prevent light from RY Tau to reach the disk at optical wavelengths (I band) while allowing light to scatter in the disk at longer wavelengths (H band). These elevated grains could be a dusty disk wind. 
    \item
    The dust mass required, in the form of a wind-launching region, to create this shadow effect in the disk is constrained to $1\times10^{-9} M_{\odot} \leq M \leq 5\times10^{-8} M_{\odot}$, corresponding to a lower limit of $\dot{M}_{\rm wind}\approx1\times10^{-8}M_{\odot}\mathrm{yr}^{-1}$ for a disk wind with a speed of $10\,\mathrm{km s}^{-1}$.
    \item 
     While an illuminate dust cavity cannot be ruled out without measurements of the gas velocity, we argue that a magnetically launched disk wind is the most likely scenario. 
    
\end{enumerate}

RY Tau shows that IMTT stars could be good candidates for observing disk winds. They are massive enough to have a radiation field able to efficiently ionize the surface layer and, in this way, allow for coupling between the disk atmosphere and wind to the magnetic field. The unique system of RY Tau shows that chances are that more winds around IMTT stars could be detected in scattered light if the star has a high accretion rate (presumably connected to a high wind mass-loss rate) and highly inclined disks (to improve the viewing geometry).


\begin{acknowledgements}
We would like to thank the anonymous referee for a fruitful discussion that lead to a significant improvement of our manuscript.
This publication is part of the project The precursors of evolved protoplanetary disks around Herbig Ae/Be stars (with project number 023.012.014) of the research programme Promotiebeurs voor Leraren which is financed by the Dutch Research Council (NWO).\\
T.Birnstiel acknowledges funding from the European Research Council (ERC) under the European Union’s Horizon 2020 research and innovation programme under grant agreement No 714769 and funding by the Deutsche Forschungsgemeinschaft (DFG, German Research Foundation) under grants 361140270, 325594231, and Germany's Excellence Strategy - EXC-2094 - 390783311.\\
C. Rab is grateful for support from the Max Planck Society and acknowledges funding by the Deutsche Forschungsgemeinschaft (DFG, German Research Foundation) - 325594231.\\
C.F. Manara acknowledges funding from the European Research Council (ERC) under the European Union’s Horizon Europe programme under grant agreement No 101039452 and wants to add that this work benefited from discussions with the ODYSSEUS team (HST AR-16129), \url{https://sites.bu.edu/odysseus/} also this project has received funding from the European Union's Horizon 2020 research and innovation programme under the Marie Sklodowska-Curie grant agreement No 823823 (DUSTBUSTERS). This work was partly funded by the Deutsche Forschungsgemeinschaft (DFG, German Research Foundation) – 325594231\\
P. Pinilla acknowledge support provided by the Alexander von Humboldt Foundation in the framework of the Sofja Kovalevskaja Award endowed by the Federal Ministry of Education and Research.\\
This work has made use of data from the European Space Agency (ESA) mission
{\it Gaia} (\url{https://www.cosmos.esa.int/gaia}), processed by the {\it Gaia}
Data Processing and Analysis Consortium (DPAC,
\url{https://www.cosmos.esa.int/web/gaia/dpac/consortium}). Funding for the DPAC
has been provided by national institutions, in particular the institutions
participating in the {\it Gaia} Multilateral Agreement.

\end{acknowledgements}

%
%

\bibliographystyle{aa}
\bibliography{RYTaubib}

\begin{appendix}

\section{Photometric data from literature and clarification of P.A. measurment}

In this appendix, we present the data used to create the SED of RY Tau. We also clarify how the angles were measured for the analysis of the observations. 

\begin{table}[h!]
\caption{Table caption for photometry}\label{tbl:photometry}
\centering
\renewcommand{\arraystretch}{1.0}
\begin{tabular}{lll}
\hline \hline
Filter & Flux & Reference \\
$ $ & $[Jy]$ & \\
\hline
$B$ & $0.126$& \citep{2015AAS...22533616H}\\
$g'$ & $0.164$ & \citep{2015AAS...22533616H}\\
$V$ & $0.266$ & \citep{2015AAS...22533616H}\\
$r'$ & $0.338$ & \citep{2015AAS...22533616H} \\[1mm]
$B$ & $0.155$& \citep{2011MNRAS.411..435B}\\
$V$ & $0.3$ & \citep{2011MNRAS.411..435B}\\[1mm]
$B$ & $0.12$ & \citep{2016MNRAS.463.4210N}\\
$V$ & $0.271$ & \citep{2016MNRAS.463.4210N}\\[1mm]
$J$ & $2.21 $ & \citep{2003yCat.2246....0C}\\
$H$ & $3.68 $ & \citep{2003yCat.2246....0C}\\
$K$ & $4.54 $ & \citep{2003yCat.2246....0C}\\[1mm]
WISE $1 $ & $6.19 $ & \citep{2012yCat.2311....0C} \\
WISE $2 $ & $9.38 $ & \citep{2012yCat.2311....0C}\\
WISE $3$ & $11.7 $ & \citep{2012yCat.2311....0C}\\
WISE $4 $ & $18.3 $ & \citep{2012yCat.2311....0C}\\[1mm]
AKARI $S9W $ & $12.3 $ & \citep{2010AA...514A...1I}\\
AKARI $L18W $ & $15.4 $ & \citep{2010AA...514A...1I}\\[1mm]
IRAC $4.5\,\mu$m & $2.42 $ & \citep{2003PASP..115..965E} \\
IRAC $5.8\,\mu$m & $4.55 $ & \citep{2003PASP..115..965E}\\
IRAC $8.0\,\mu$m & $6.56 $ & \citep{2003PASP..115..965E}\\[1mm]
IRAS $12 $ & $17.5 $ & \citep{2015AC....10...99A}\\
IRAS $25 $ & $26.1 $ & \citep{2015AC....10...99A}\\
IRAS $60 $ & $15.3 $ & \citep{2015AC....10...99A}\\
IRAS $100 $ & $13.6 $ & \citep{2015AC....10...99A}\\[1mm]
SMA $1.3$\,mm & $0.192 $ & \citep{2013ApJ...771..129A}\\
SMA $886$\,$\mu$m & $0.499 $ & \citep{2013ApJ...771..129A}\\[1mm]
SCUBA $1.3$\,mm  & $0.229$ & \citep{2013ApJ...773..168M}\\
SCUBA $886$\,$\mu$m & $0.56$ & \citep{2013ApJ...773..168M}\\[1mm]
VLA $6.7$\,cm  & $1.84\times10^{-4}$ & \citep{2015ApJ...801...91D}\\
VLA $4$\,cm  & $2.59\times10^{-4}$ & \citep{2015ApJ...801...91D}\\[1mm]
PACS $70 $ & $14.0 $ & \citep{2017ApJ...849...63R}\\
PACS $160 $ & $9.0 $ & \citep{2017ApJ...849...63R}\\[1mm]
SPIRE $250 $ & $5.0 $ & \citep{2017ApJ...849...63R}\\
SPIRE $350$ & $3.2 $ & \citep{2017ApJ...849...63R}\\
SPIRE $500$ & $1.8 $ & \citep{2017ApJ...849...63R}\\

\hline
\end{tabular}
\end{table}

   \begin{figure}[b!]
        \centering
        \includegraphics[width=0.4\textwidth]{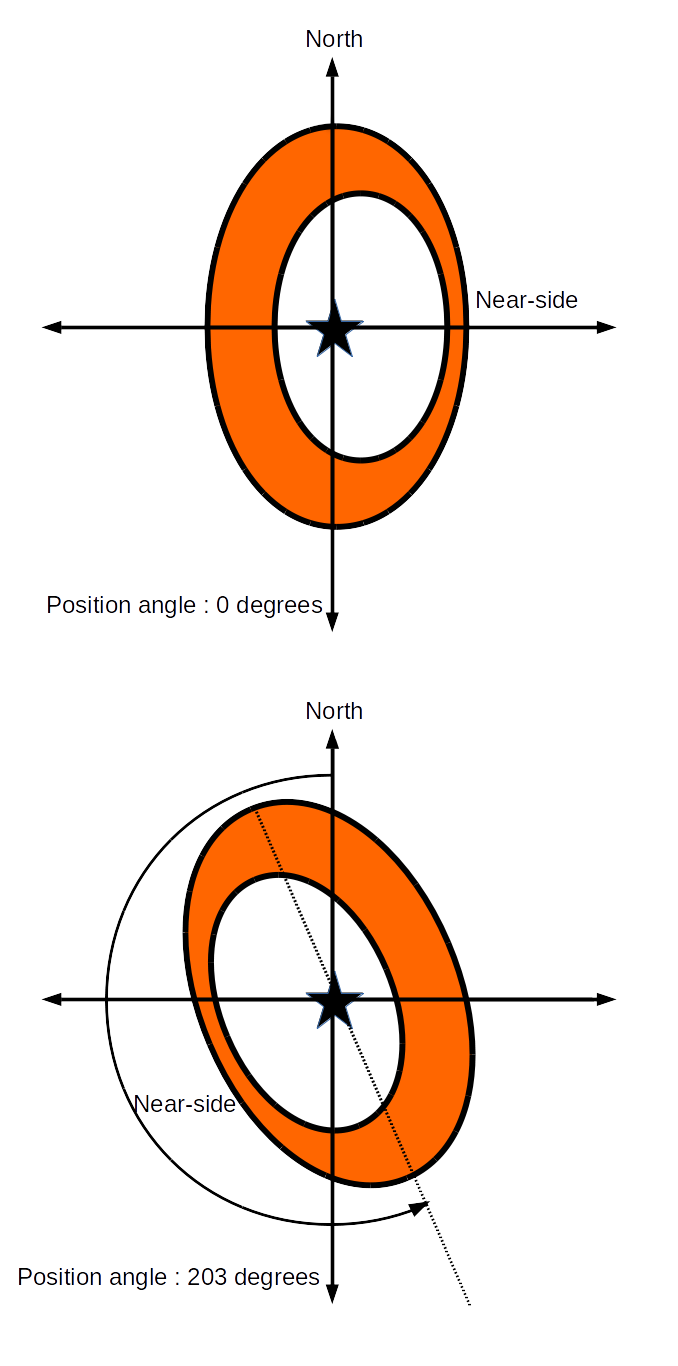}
        \caption{Position angle measured anticlockwise from north where P.A.=0$^{\circ}$ is when the near-side of the disk was located due west (top image). In the case of RY Tau, the near side of the disk is located in the south-east, thus P.A.= 203$^{\circ}$, and measured as shown by the lower image.}
        \label{fig:orientation}
    \end{figure}

\end{appendix}

\end{document}